\documentstyle[twoside,epsf,psfig,times]{article}
\def\beqa{\begin{eqnarray}}
\def\eeqa{\end{eqnarray}}
\def\beq{\begin{equation}}
\def\eeq{\end{equation}}
\catcode`\@=11
\long\def\@makefntext#1{
\protect\noindent \hbox to 3.2pt {\hskip-.9pt  
$^{{\eightrm\@thefnmark}}$\hfil}#1\hfill}               

\def\@makefnmark{\hbox to 0pt{$^{\@thefnmark}$\hss}}    

\def\ps@myheadings{\let\@mkboth\@gobbletwo
\def\@oddhead{\hbox{}
\rightmark\hfil\eightrm\thepage}   
\def\@oddfoot{}\def\@evenhead{\eightrm\thepage\hfil
\leftmark\hbox{}}\def\@evenfoot{}
\def\sectionmark##1{}\def\subsectionmark##1{}}

\oddsidemargin=\evensidemargin
\newcounter{sectionc}\newcounter{subsectionc}\newcounter{subsubsectionc}
\renewcommand{\section}[1] {\vspace{12pt}\addtocounter{sectionc}{1} 
\setcounter{subsectionc}{0}\setcounter{subsubsectionc}{0}\noindent 
        {\tenbf\thesectionc. #1}\par\vspace{5pt}}
\renewcommand{\subsection}[1] {\vspace{12pt}\addtocounter{subsectionc}{1} 
\setcounter{subsubsectionc}{0}\noindent 
{\bf\thesectionc.\thesubsectionc. {\kern1pt \bfit #1}}\par\vspace{5pt}}
\renewcommand{\subsubsection}[1] {\vspace{12pt}\addtocounter{subsubsectionc}{1}
        \noindent{\tenrm\thesectionc.\thesubsectionc.\thesubsubsectionc.
        {\kern1pt \tenit #1}}\par\vspace{5pt}}
\newcommand{\nonumsection}[1] {\vspace{12pt}\noindent{\tenbf #1}
        \par\vspace{5pt}}

\newcounter{appendixc}
\newcounter{subappendixc}[appendixc]
\newcounter{subsubappendixc}[subappendixc]
\renewcommand{\thesubappendixc}{\Alph{appendixc}.\arabic{subappendixc}}
\renewcommand{\thesubsubappendixc}
        {\Alph{appendixc}.\arabic{subappendixc}.\arabic{subsubappendixc}}

\renewcommand{\appendix}[1] {\vspace{12pt}
        \refstepcounter{appendixc}
        \setcounter{figure}{0}
        \setcounter{table}{0}
        \setcounter{lemma}{0}
        \setcounter{theorem}{0}
        \setcounter{corollary}{0}
        \setcounter{definition}{0}
        \setcounter{equation}{0}
        \renewcommand{\thefigure}{\Alph{appendixc}.\arabic{figure}}
        \renewcommand{\thetable}{\Alph{appendixc}.\arabic{table}}
        \renewcommand{\theappendixc}{\Alph{appendixc}}
        \renewcommand{\thelemma}{\Alph{appendixc}.\arabic{lemma}}
        \renewcommand{\thetheorem}{\Alph{appendixc}.\arabic{theorem}}
        \renewcommand{\thedefinition}{\Alph{appendixc}.\arabic{definition}}
        \renewcommand{\thecorollary}{\Alph{appendixc}.\arabic{corollary}}
        \renewcommand{\theequation}{\Alph{appendixc}.\arabic{equation}}
        \noindent{\tenbf Appendix \theappendixc #1}\par\vspace{5pt}}
\newcommand{\subappendix}[1] {\vspace{12pt}
        \refstepcounter{subappendixc}
        \noindent{\bf Appendix \thesubappendixc. {\kern1pt \bfit #1}}
        \par\vspace{5pt}}
\newcommand{\subsubappendix}[1] {\vspace{12pt}
        \refstepcounter{subsubappendixc}
        \noindent{\rm Appendix \thesubsubappendixc. {\kern1pt \tenit #1}}
        \par\vspace{5pt}}

\newcommand{\textlineskip}{\baselineskip=13pt}
\newcommand{\smalllineskip}{\baselineskip=10pt}

\newcommand{\copyrightheading}[1]
        {\vspace*{-2.5cm}\smalllineskip{\flushleft
        {\footnotesize International Journal of Modern Physics D, #1}\\
        {\footnotesize \copyright\kern2pt World Scientific Publishing
         Company}\\
         }}

\newcommand{\publisher}[2]{{\begin{center}\footnotesize\smalllineskip 
        Received #1\\
        Revised #2
        \end{center}
        }}

\def\abstracts#1#2#3{{
        \centering{\begin{minipage}{4.5in}\footnotesize\baselineskip=10pt
        \parindent=0pt #1\par 
        \parindent=15pt #2\par
        \parindent=15pt #3
        \end{minipage}}\par}} 

\renewenvironment{thebibliography}[1]
        {\frenchspacing
         \ninerm\baselineskip=11pt
         \begin{list}{\arabic{enumi}.}
        {\usecounter{enumi}\setlength{\parsep}{0pt}     
         \setlength{\leftmargin 12.7pt}{\rightmargin 0pt}
         \setlength{\itemsep}{0pt} \settowidth
        {\labelwidth}{#1.}\sloppy}}{\end{list}}

\newcounter{itemlistc}
\newcounter{romanlistc}
\newcounter{alphlistc}
\newcounter{arabiclistc}

\newcommand{\fcaption}[1]{
        \refstepcounter{figure}
        \setbox\@tempboxa = \hbox{\footnotesize Fig.~\thefigure. #1}
        \ifdim \wd\@tempboxa > 5in
           {\begin{center}
        \parbox{5in}{\footnotesize\smalllineskip Fig.~\thefigure. #1}
            \end{center}}
        \else
             {\begin{center}
             {\footnotesize Fig.~\thefigure. #1}
              \end{center}}
        \fi}

\newcommand{\tcaption}[1]{
        \refstepcounter{table}
        \setbox\@tempboxa = \hbox{\footnotesize Table~\thetable. #1}
        \ifdim \wd\@tempboxa > 5in
           {\begin{center}
        \parbox{5in}{\footnotesize\smalllineskip Table~\thetable. #1}
            \end{center}}
        \else
             {\begin{center}
             {\footnotesize Table~\thetable. #1}
              \end{center}}
        \fi}

\def\@citex[#1]#2{\if@filesw\immediate\write\@auxout
        {\string\citation{#2}}\fi
\def\@citea{}\@cite{\@for\@citeb:=#2\do
        {\@citea\def\@citea{,}\@ifundefined
        {b@\@citeb}{{\bf ?}\@warning
        {Citation `\@citeb' on page \thepage \space undefined}}
        {\csname b@\@citeb\endcsname}}}{#1}}

\newif\if@cghi
\def\cite{\@cghitrue\@ifnextchar [{\@tempswatrue
        \@citex}{\@tempswafalse\@citex[]}}
\def\citelow{\@cghifalse\@ifnextchar [{\@tempswatrue
        \@citex}{\@tempswafalse\@citex[]}}
\def\@cite#1#2{{$\null^{#1}$\if@tempswa\typeout
        {IJCGA warning: optional citation argument 
        ignored: `#2'} \fi}}

\def\pmb#1{\setbox0=\hbox{#1}
        \kern-.025em\copy0\kern-\wd0
        \kern.05em\copy0\kern-\wd0
        \kern-.025em\raise.0433em\box0}

\def\fnt#1#2{\footnotetext{\kern-.3em
        {$^{\mbox{\scriptsize #1}}$}{#2}}}

\def\fpage#1{\begingroup
\voffset=.3in
\thispagestyle{empty}\begin{table}[b]\centerline{\footnotesize #1}
        \end{table}\endgroup}

\def\runninghead#1#2{\pagestyle{myheadings}
\markboth{{\protect\footnotesize\it{\quad #1}}\hfill}
{\hfill{\protect\footnotesize\it{#2\quad}}}}
\font\tenrm=cmr10
\font\tenit=cmti10 
\font\tenbf=cmbx10
\font\bfit=cmbxti10 at 10pt
\font\ninerm=cmr9

\font\eightrm=cmr8

\def\qed{\hbox{${\vcenter{\vbox{                  
   \hrule height 0.4pt\hbox{\vrule width 0.4pt height 6pt
   \kern5pt\vrule width 0.4pt}\hrule height 0.4pt}}}$}}

\begin{document}
\setlength{\textheight}{7.7truein}    

\runninghead{Bulk Viscous Cosmological Models in Barber's Second Self Creation Theory } 
{A. Pradhan and H. R. Pandey}

\normalsize\textlineskip
\thispagestyle{empty}
\setcounter{page}{1}

\copyrightheading{}             {Vol.~0, No.~0 (2002) 000--000}

\vspace*{0.88truein}

\fpage{1}

\centerline{\bf BULK VISCOUS COSMOLOGICAL MODELS IN}
\vspace*{0.035truein}
\centerline{\bf BARBER'S SECOND SELF CREATION THEORY}
\vspace*{0.37truein}
\centerline{\footnotesize ANIRUDH PRADHAN\footnote{E-mail: acpradhan@yahoo.com,
apradhan@mri.ernet.in (Corresponding Author)}}
\vspace*{0.015truein}
\centerline{\footnotesize\it Department of Mathematics, Hindu Post-graduate College,}
\baselineskip=10pt
\centerline{\footnotesize\it Zamania, Ghazipur 232 331, India}
\vspace*{10pt}
\centerline{\footnotesize HARE RAM PANDEY}
\vspace*{0.015truein}
\centerline{\footnotesize\it Department of Mathematics, R. S. K. I. College,}
\baselineskip=10pt
\centerline{\footnotesize\it Dubahar, Ballia 277 405, India}
\baselineskip=10pt
\vspace*{0.225truein}
\publisher{(received date)}{(revised date)}
\vspace*{0.21truein}
\abstracts{Barber's second self creation theory with bulk viscous fluid source
for an LRS Bianchi type-I metric is considered by using deceleration 
parameter to be constant where the metric potentials are taken as function 
of $x$ and $t$. The coefficient of bulk viscosity is assumed to be a power
function of the mass density. Some physical and geometrical features of the
models are discussed.}{}{}


\vspace*{1pt}\textlineskip      
\section{Introduction}
\vspace*{-0.5pt}
\noindent
Several modification of Einstein's general relativity have been proposed
and extensively studied so far by many cosmologists to unify gravitation 
and many other effects in the universe. Barber \cite{ref1} has produced two
continuous self-creation theories by modifying the Brans-Dicke theory and
general relativity. The modified theories create the universe out of self-contained 
gravitational and matter fields. Brans-Dicke \cite{ref2} theory 
develops Mach's principle in a relativistic framework by assuming 
interaction of inertial masses of fundamental particles with some cosmic 
scalar field coupled with the large scale distribution of matter in motion. 
However, Barber \cite{ref1} has included continuous creation of matter in these 
theories. The universe is seen to be created out of self-contained gravitational,
scalar and matter fields. However, the solution of the one-body problems
reveal unsatisfactory characteristics of the first theory and in 
particular the principle of equivalence is severely violated. The second 
theory retains the attractive features of the first theory and overcomes 
previous objections. These modified theories create the universe out of 
self-contained gravitational and matter fields. Recently, Brans \cite{ref3} has 
also pointed out that Barber's first theory is in disagreement with 
experiment as well as inconsistent, in general, since the equivalence 
principle is violated. Barber's second theory is a modification of general 
relativity to a variable G-theory. The second is an adaptation of general 
relativity to include continuous creation and is within the observational 
ambit. In this theory the scalar fields do not directly gravitate, but simply 
divide the matter tensor, acting as a reciprocal gravitational constant. 
It is postulated that this scalar field couples with the trade of energy 
momentum tensor. In view of the consistency of Barber's second theory of 
gravitation, we intend to investigate some of the aspects of this theory in 
this paper.
\newline
\par
Several cosmologists have studied various aspects of Robertson-Walker
model in Barber's second self-creation cosmology with perfect fluid
satisfying the equation of state $p = (\gamma -1)\rho$, where $1\leq
\gamma\leq 2$. Pimentel \cite{ref4} and Soleng \cite{ref5} have discussed the 
Robertson-Walker solutions in Barber's second self-creation theory of 
gravitation by using power law relation between the expansion factor of 
the universe and the scalar field. Carvalho \cite{ref6} studied a homogeneous 
and isotropic model of the early universe in which parameter gamma of 
`gamma law' equation of state varies continuously with cosmological time and 
presented a unified description of early universe for inflationary period 
and radiation-dominating era. Singh, \cite{ref7} Reddy, \cite{ref8,ref9} and 
Reddy et al \cite{ref10} have presented Bianchi type space-times solutions 
in Barber's second theory of gravitation. Reddy and Venkateswarlu \cite{ref11} 
present Bianchi type $-VI_0$ cosmological solutions in Barber's second theory 
of gravitation both, in vacuum as well as in the presence of perfect fluid 
with pressure equal to energy density. Shanthi and Rao \cite{ref12} studied 
Bianchi type II and III space-times in second theory of gravitation, both 
in vacuum as well as in presence of stiff-fluid. Recently, Shri Ram and 
Singh \cite{ref13} have discussed spatially homogeneous and isotropic R-W model 
of the universe in Barber's second self-creation theory of gravitation in the 
presence of perfect fluid by using `gamma-law' equation of state. Barber's 
second self creation theory with perfect fluid source for an LRS Bianchi type-I 
metric is considered recently by Pradhan and Vishwakarma \cite{ref14} by using 
deceleration parameter to be constant where the metric potentials are taken as 
function of $x$ and $t$.    
\newline
\par
Most studies in cosmology involve a perfect fluid. However, observed phenomena 
such as the large entropy per baryon and the remarkable degree of isotropy
of the cosmic microwave background radiation, suggest that we should analyse 
dissipative effects in cosmology. Further, there are several processes which 
are expected to give rise to viscous effects. These are the decoupling of 
neutrinos during the radiation era and the decoupling of matter and radiation 
during the recombination era. Bulk viscosity is associated with the GUT phase 
transition and string creation. It is known that the introduction of bulk 
viscosity can avoid the big bang singularity. For further discussions, see 
the references in Pradhan et al.{\cite{ref15}$^-$\cite{ref18}} 
\newline
\par
For simplification and description of the large scale behaviour of the actual
universe, locally rotationally symmetric [henceforth referred as LRS] Bianchi I 
spacetime have widely studied.{\cite{ref19}$^-$\cite{ref24}} When the Bianchi I
spacetime expands equally in two spatial directions it is called locally rotationally
symmetric. These kinds of models are interesting because Lidsey \cite{ref25} showed
that they are equivalent to a flat (FRW) universe with a self-interacting 
scalar field and a free massless scalar field, but produced no explicit example.
Some explicit solutions were pointed out in Refs. \cite{ref26,ref27} Motivated 
by these above arguments, in this paper, we have investigated spatially flat and 
non-flat bulk viscous LRS Bianchi type-I cosmological models in Barber's second 
theory of gravitation.
\newline
\par

\section{Field Equations}
\noindent
We consider an LRS Bianchi type-I spacetime
\begin{equation}
\label{eq1}
ds^2 = dt^2 -  A^2dx^2 - B^2 (dy^2 + dz^2)
\end{equation}
\noindent 
where $A=A(x,t), B=B(x,t)$. The field equations in Barber's second self-creation 
theory \cite{ref1} are
\begin{equation}
\label{eq2}
R_{ij} - \frac{1}{2} g_{ij} R  = - \frac{8 \pi  T_{ij}}{\phi} 
\end{equation}  
and
\begin{equation}
\label{eq3}
\Box \phi = 4 \pi \lambda T
\end{equation}
where $\Box \phi \equiv \phi^{;k}_{;k}$ is the invariant d'Alemberian
and $T$ is the trace of the energy momentum tensor. $\lambda$ is coupling
constant to be determined from the experiment $ (|\lambda| \leq 0.1)$.
In the limit $\lambda \rightarrow 0$ this theory approaches the standard
general relativity theory in every respect and $G = \frac{1}{\phi}$.
\noindent 
The stress-energy tensor in the presence of bulk stress has the form
\begin{equation}
\label{eq4}
T_{ij} = (\bar{p} +\rho) u_i u_j - \bar{p} g_{ij}
\end{equation}
\noindent where
\begin{equation}
\label{eq5}
\bar{p} = p - \xi u^{i}_{;i} ~~~ and ~~~ u^{i} u_{i} 
\end{equation}
\noindent 
Here $\rho$, $p$, $\xi$ and $u$ are, respectively, the energy density,
isotropic pressure, bulk viscous coefficient and four-velocity vector of
the distribution. Corresponding to metric (1), the four velocity 
vector $u_{i}$ satisfies the equation
\begin{equation}
\label{eq6}
g_{ij} u^i u^j = 1
\end{equation}
\noindent 
The Bianchi identities in contravariant form applied to Eq.  (\ref{eq2}) are
\begin{equation}
\label{eq7}
W_{;i}T^{ij} + W T^{ij}_{;i} = 0
\end{equation}
where $ W = - 8\pi\phi^{-1}$ and  in general relativity $W = - 8\pi G$.
A comma and a semicolon denote ordinary and covariant differentiation, respectively. 
In a comoving coordinate system, the surviving components
of the field Eqs. (\ref{eq2})-(\ref{eq7}) for metric (1) are
\begin{equation}
\label{eq8}
\frac{2\ddot B}{B} + \frac{\dot B^2}{B^2} - \frac{B'^2}{A^2 B^2}
 = - 8\pi \phi^{-1} \bar{p}
\end{equation}
\begin{equation}
\label{eq9}
\dot B' - \frac{B' \dot A}{A} = 0 
\end{equation}
\begin{equation}
\label{eq10}
\frac{\ddot A}{A} + \frac{\ddot B}{B} + \frac{\dot A \dot B}
{AB} - \frac{B''}{A^2 B} + \frac{A' B'}{A^3 B} = -8\pi\phi^{-1}\bar{p}
\end{equation}
\begin{equation}
\label{eq11}
\frac{2B''}{A^2 B} - \frac{2 A' B'} {A^3 B} + \frac{B'^2}
{A^2 B^2} - \frac{2 \dot A \dot B}{AB} - \frac{\dot B^2}{B^2}
 = 8\pi\phi^{-1}\rho
\end{equation}
\begin{equation}
\label{eq12}
\ddot \phi +\frac{\dot A \dot \phi}{A} + \frac{2 \dot B \dot \phi}{B}
+\frac{A'\phi'}{A^3} - \frac{2 B' \phi'}{A^2 B} -\frac{\phi''}{A} =
(\rho - 3\bar{p}) (\frac{8\pi \lambda}{3})
\end{equation}
Here and in following expressions a prime and a dot indicate partial 
differentiation with respect to $x$ and $t$ respectively. \\
\section{Solution of the Field Equations and Discussion}
\noindent  In this section, we review the solutions obtained by Pradhan
and Vishwakarma. \cite{ref14} Eq. (\ref{eq9}), after integration, yields
\begin{equation}
\label{eq13}
A = \frac{B'}{\l}
\end{equation}
\noindent  where $\l$ is an arbitrary function of $x$ \\
\noindent Eqs. (\ref{eq8}) and (\ref{eq10}), with the use of Eq. (\ref{eq13}),
reduce to
\begin{equation}
\label{eq14}
\frac{B}{B'} \frac{d}{dx}\left(\frac{\ddot B}{B}\right) + \frac{\dot B}{B'}
\frac{d}{dt} \left(\frac{B'}{B}\right) + \frac{\l^2}{B^2}
\left(1 - \frac{B \l'}{B'\l}\right) = 0
\end{equation}
\noindent If we assume $\frac{B'}{B}$ as a function of x alone, then  $A$ and $B$
are separable in $x$ and $t$, Eq. (\ref{eq14}) gives after integration  
\begin{equation}
\label{eq15}
B = \l S(t)
\end{equation}
\noindent  where $S(t)$ is an arbitrary function of $t$.
\noindent  With the help of Eq. (\ref{eq15}), Eq. (\ref{eq13}) becomes
\begin{equation}
\label{eq16}
A = \frac{\l'}{\l} S
\end{equation}
\noindent  Now the metric (1) takes the form
\begin{equation}
\label{eq17}
ds^2 = dt^2 - S^2(t) [dX^2 + e^{2X} (dy^2 + dz^2)]
\end{equation}
\noindent  where $X = \ln \l $.
\noindent  With the use of Eqs. (\ref{eq15}) and (\ref{eq16}), Eqs. (\ref{eq8}),
(\ref{eq11}) and (\ref{eq12}) yield
\begin{equation}
\label{eq18}
\frac{2 \ddot S}{S} + \frac{\dot S^2}{S^2} - \frac{1}{S^2} =
-8\pi \phi^{-1} \bar{p}
\end{equation}
\begin{equation}
\label{eq19} 
\frac{3}{S^2} - \frac{3 \dot S^2}{S^2} = 8\pi \phi^{-1} \rho
\end{equation}
\begin{equation}
\label{eq20}
{\ddot \phi} + \frac{3\dot \phi\dot S}{S} -\frac{3\phi' \l}{\l' S^2}
- \frac{\l^2}{\l' S^2} \frac{d}{dx}\left(\frac{\phi'}{\l'}\right)
= \frac{8\pi \lambda}{3}(\rho - 3\bar{p})
\end{equation}
\noindent  For the sake of simplicity, if we assume $\phi$ to be a function of
$t$ only, then Eq. (\ref{eq20}) with the use of Eqs. (\ref{eq18}) and (\ref{eq19})
reduces to
\begin{equation}
\label{eq21}
\frac{\ddot\phi}{\phi} + \frac{3 {\dot \phi} \dot S}{\phi S} -
\frac{2 \lambda \ddot S}{S} = 0
\end{equation}
\noindent  The function $S(t)$ remains undetermined. To obtain its explicit 
dependence on $t$, one may have to introduce additional assumptions.
In the following, we assume the deceleration parameter to be constant
to achieve this objective i.e.
\begin{equation}
\label{eq22}
q = - \frac{S \ddot S}{{\dot S}^2} = - \left(\frac{\dot H + H^2}{H^2}\right) 
= b \mbox{(constant)},
\end{equation}
\noindent  where, $H = \dot S/S$ is the Hubble parameter. The above equation may be 
rewritten as 
\begin{equation}
\label{eq23}
\frac{\ddot S}{S} + b \frac{{\dot S}^2}{S^2}=0
\end{equation}
\noindent  On integration  Eq. (\ref{eq23}) gives the exact solution
\begin{equation}
\label{eq24}
S(t) = \left\{ \begin{array}{ll}
              [a(t - t_0)]^{\frac{1}{(1+b)}}  & \mbox{when $b\neq-1$}\\
               m_1 e^{m_2t}                        & \mbox{when $b=-1$}
              \end{array} \right. 
\end{equation}
\noindent  where $a, m_1$ and $m_2$ are constants of integration and the constant
$t_0$ means the freedom of choosing the time origin.
\noindent Using Eq. (\ref{eq22}) in Eqs. (\ref{eq18}), (\ref{eq19}) and (\ref{eq21}) lead to
\begin{equation}
\label{eq25}
8\pi\phi^{-1}\bar{p} = \frac{1}{S^2} + ( 2 b -1)H^2
\end{equation}
\begin{equation}
\label{eq26}
8\pi\phi^{-1}\rho = 3 (\frac{1}{S^2} - H^2)
\end{equation}
\begin{equation}
\label{eq27}
\frac{\ddot\phi}{\phi} + \frac{3\dot\phi H}{\phi} + 2\lambda b  H^2 = 0
\end{equation}

\subsection{Non-flat Models}
\noindent {\bf Case (i):} $b \neq -1$. For singular models since $S(0)=0$, Eq. 
(\ref{eq24}) leads to
\begin{equation}
\label{eq28}
S=m t^{ \frac{1}{(1+b)}}
\end{equation}
\noindent  Using Eq. (\ref{eq28}) in Eqs. (\ref{eq27}), (\ref{eq25}) and (\ref{eq26}) yield
\begin{equation}
\label{eq29}
\phi= m^{- \frac{3}{2}} t^{- \frac{3}{2(1+b)}} \left[ c_1 e^{ \frac{t}{2} 
( 1 + \sqrt{1-4k_1})} + c_2 e^{ \frac{t}{2} ( 1 - \sqrt{1-4k_1})} \right]
\end{equation}
\begin{equation}
\label{eq30}
8 \pi (p - \xi u^{i}_{;i}) = \phi \left[ \frac{1}{m^2 t^{\frac{2}{(1+b)}}} + 
\frac{(2b-1)}{{(1+b)}^2 t^2} \right]
\end{equation}
\begin{equation}
\label{eq31}
8 \pi \rho =3 \phi \left[ \frac{1}{m^2 t^{\frac{2}{(1+b)}}} -
\frac{1}{{(1+b)}^2 t^2} \right]
\end{equation}
\noindent where
\begin{equation}
\label{eq32}
k_1= \frac{[2b(4 \lambda +3)-3 ]}{4{(1+b)}^2}
\end{equation}
\noindent  and $c_1$ and $c_2$ are arbitrary constants of integration. Therefore,
the geometry of the universe, in this case, is described by the line-element
\begin{equation}
\label{eq33}
d s^2 = d t^2 -m^2t^{\frac{2}{(1+b)}} [ d X^2 + e^{2X} ( d y^2 + d z^2)]
\end{equation}
\noindent  Here the Barber's scalar function $\phi$ is given by Eq. (\ref{eq29}) and the 
corresponding physical parameters $p$ and $\rho$ are given by the Eqs. (\ref{eq30})
and (\ref{eq31}) respectively.\\
\noindent  Thus, for given $\xi(t)$, we can solve the pressure $p$. It is standard to
assume \cite{ref28,ref29} the following ad-hoc law
\begin{equation}
\label{eq34}
\xi(t) = \xi_{0} \rho^{n}
\end{equation}
\noindent If $n = 1$, Eq. (\ref{eq34}) may correspond to a radiative fluid, whereas 
$n = \frac{3}{2}$ may correspond to a string dominated universe. \cite{ref30}
However, more realistic models \cite{ref31} are based on $n$ lying in the region 
$0 \leq n \leq \frac{1}{2}$.\\
\bigskip
\noindent {\bf Model I $(\xi = \xi_0)$}\\
\noindent  In this case, we assume $n = 0$. Further from Eq. (\ref{eq34}), we obtain $\xi$ = 
$\xi_O$ = constant and hence Eq.(\ref{eq30}) becomes
\begin{equation}
\label{eq35}
p = \frac{\phi}{8 \pi t^2}[\frac{1}{m^2 t^{\frac{1}{(1 + b)}}} + \frac{(2b -1)}{(1 + b)^2}]
- \frac{2\xi_0}{(1 + b)t}
\end{equation}
\bigskip
\noindent {\bf Model II $(\xi = \xi_0 \rho)$}\\
\noindent  In this case, we assume $n = 1$. Further from Eq. (\ref{eq34}), we obtain $\xi$ = 
$\xi_O \rho$ and hence Eq. (\ref{eq30}) reduces to
\begin{equation}
\label{eq36}
p = \frac{\phi}{8 \pi t^2}[\frac{1}{m^2 t^{\frac{1}{(1 + b)}}} + \frac{(2b - 1)}
{(1 + b)^2} - \frac{9 \xi}{(1 + b)t}\{\frac{1}{m^2 t^\frac{1}{(1 + b)}} - \frac{1}{(1 + b)^2}\}]
\end{equation}
\noindent  From Eq. (\ref{eq31}), we observe that the energy density $\rho(t)$ is decreasing
function of time. As $t$ tends to infinity, energy density will vanish. From Eq. (\ref{eq29}) , 
it is observed that Barber's scalar function $\phi$ decreases as time increases 
and will vanish when $t$ turns to infinity. The energy conditions \cite{ref32,ref33} are 
satisfied provided $ m^2 <1$ and $ b = 0$ with positive $c_1$ and $c_2$.\\
\bigskip
\noindent {\bf Physical behaviour of the models:}
In case of a non-flat model when $b \neq -1$, the Ricci scalar becomes
\begin{equation}
\label{eq37}
R= \frac{1}{m^2 t^{\frac{2}{(1+b)}}} - {\frac{(1-b)t}{(1+b)}} 
\end{equation}
\noindent It is observed from Eq. (\ref{eq37}) that when $ t \rightarrow 0$;
(i) $ R \rightarrow \infty$ if $b=0$, and
(ii) $ R \rightarrow \infty$ if $b \geq 1$.  
The Eq. (\ref{eq37}) also suggests that when $ t \rightarrow \infty$;
$ R \rightarrow 0$ if $b \geq 0$. 
\noindent The scalars of the expansion and shear are given by 
\begin{equation}
\label{eq38}
\theta= {\frac{3}{(1+b)t}}~ , ~~\sigma=0
\end{equation}
\noindent The model has singularity at $t=0$. At $ t \rightarrow \infty$,
the expansion ceases. Here, $ \frac{\sigma}{\theta}=0$, which confirms the 
isotropic nature of the space-time which we have already obtained in 
Eq. (\ref{eq33}).\\
\bigskip
\noindent {\bf Case (ii):} $b= -1$. In this case using Eq. (\ref{eq24}) in Eqs. (\ref{eq27}), 
(\ref{eq25}) and (\ref{eq26}) give 
\begin{equation}
\label{eq39}
\phi = {m_1}^{- \frac{3}{2}} e^{-\frac{3m_2 t}{2}} 
\left[c_3 e^{{\sqrt k_2}t} + c_4 e^{-{\sqrt k_2}t} \right]
\end{equation}
\begin{equation}
\label{eq40}
8 \pi(p + 3m_2 \xi) = \phi \left[ \frac{1}{{m_1}^2 e^{2 m_2 t}} - 3 {m_2}^2 \right]
\end{equation}
\begin{equation}
\label{eq41}
8 \pi \rho =3 \phi \left[ \frac{1}{{m_1}^2 e^{2 m_2 t}} -  {m_2}^2 \right]
\end{equation}
\noindent where
\begin{equation}
\label{eq42}
k_2= {m_2}^2 \left[ 2 \lambda + \frac{9}{4} \right]
\end{equation}
\noindent and $c_3$ and $c_4$ are arbitrary constants of integration.
\noindent Therefore, the geometry of the universe, in this case, is 
described by the line-element
\begin{equation}
\label{eq43}
d s^2= dt^2 - {m_1}^2 e^{2m_2 t}[ d X^2 + e^{2X} ( d y^2 + d z^2)]
\end{equation}
\noindent In this case, the Barber's scalar function $\phi$ is given by the 
Eq. (\ref{eq39}) and the corresponding physical parameters $p$ and $\rho$ are given by 
Eqs.  (\ref{eq40}) and (\ref{eq41}) respectively. For $m_2 = 0$, the model reduces to 
a static radiating model with constant density and pressure.  \\
\bigskip
\noindent {\bf Model I $(\xi = \xi_0)$}\\
\noindent In this case, we assume $n = 0$. Further from Eq. (\ref{eq34}), we obtain 
$\xi$ = $\xi_O $= constant and hence Eq. (\ref{eq40}) reduces to
\begin{equation}
\label{eq44}
p = \frac{\phi}{8\pi}[\frac{1}{m_1^2 e^{2m_2 t}} - 3m_2^2] -3m_2 \xi_0
\end{equation}
\bigskip
\noindent {\bf Model II $(\xi = \xi_0 \rho)$}\\
\noindent In this case, we assume $n = 1$. Further from Eq. (\ref{eq34}), we obtain 
$\xi$ = $\xi_O \rho$ and hence Eq. (\ref{eq40}) reduces to
\begin{equation}
\label{eq45}
p = \frac{\phi}{8\pi}[\frac{1}{m_1^2 e^{2m_2 t}} - 3m_2^2 - 9m_2 \xi_0 \{\frac{1}
{m_1^2 e^{2m_2t}} -m_2^2\}]
\end{equation}
\bigskip
\noindent {\bf Physical behaviour of the model:}
The Ricci scalar $R$ is 
\begin{equation}
\label{eq46}
R= 2 {m_2}^2 - \frac{1}{{m_1}^2 e^{2 m_2 t}}
\end{equation}
\noindent It is easily observed from Eq. (\ref{eq46}) that (i) when 
$t \rightarrow 0,~~ R \rightarrow (2 {m_2}^2 - \frac{1}{{m_1}^2})$, and
(ii) when $t \rightarrow \infty, ~ R \rightarrow 2 {m_2}^2$. The expansion 
and shear scalars are
\begin{equation}
\label{eq47}
\theta=3 m_2, ~ ~ \sigma=0
\end{equation}
\noindent The model represents an uniform expansion as can be seen from Eq. (\ref{eq47}).
The flow of the fluid is geodetic as the acceleration vector $f_i=(0,0,0,0)$. \\

\subsection{Flat Models}
Motivated by the results of the BOOMERANG experiment on Cosmic Microwave 
Background Radiation, \cite{ref34} we wish to study a spatially flat bulk viscous 
cosmological models.\\
\noindent The condition for the flat model is obtained as 
\begin{equation}
\label{eq48}
\frac{1}{S^2}=(1-b) H^2
\end{equation}
\noindent Using Eq. (\ref{eq48}), Eqs. (\ref{eq25}) and (\ref{eq26}) reduce to
\begin{equation}
\label{eq49}
8 \pi (p - \xi u^i_{;i}) = \phi b H^2
\end{equation}
\begin{equation}
\label{eq50}
8 \pi \rho = -3 \phi b H^2~, ~~ \mbox{where}~ b<0
\end{equation}
\noindent In this case the model has $p<0$ which might describe the very early 
epoch of galaxy formation from the process of matter condensation.\\
\bigskip
\noindent {\bf Case (i):} $b \neq -1$. Using Eq. (\ref{eq28}) in Eqs. (\ref{eq49}) 
and (\ref{eq50}) yield
\begin{equation}
\label{eq51}
8 \pi (p - \frac{3\xi}{(1 + b)^2 t^2}) = \frac{\phi b}{{(1+b)}^2 t^2}
\end{equation}
\begin{equation}
\label{eq52}
8 \pi \rho = - \frac{3 \phi b }{{(1+b)}^2 t^2}~, ~~\mbox{with} ~ b<0 
\end{equation}
\noindent where $\phi$ is already given by Eq. (\ref{eq29}). Here, the model 
describes the early phase of evolution as mentioned earlier.\\
\bigskip
\noindent {\bf Model I $(\xi = \xi_0)$}\\
\noindent In this case, we assume $n = 0$. Further from Eq. (\ref{eq34}), we obtain 
$\xi$ = $\xi_O $= constant and hence Eq. (\ref{eq51}) reduces to
\begin{equation}
\label{eq53}
p = \frac{1}{(1 + b)t}[\frac{\phi b}{8\pi(1 + b)t} + 3\xi_0]
\end{equation}
\bigskip
\noindent {\bf Model II $(\xi = \xi_0 \rho)$}\\
\noindent In this case, we assume $n = 1$. Further from Eq. (\ref{eq34}), we obtain 
$\xi$ = $\xi_O \rho$ and hence Eq.(\ref{eq51}) reduces to
\begin{equation}
\label{eq54}
p = \frac{\phi b}{8\pi(1 +b)^2 t^2}[1 - \frac{9\xi_0}{(1 +b)t}]
\end{equation}
\bigskip
\noindent {\bf Case (ii):} $b= -1$. In this case using Eq. (\ref{eq24}) in Eqs.  (\ref{eq49})
and (\ref{eq50}) yield
\begin{equation}
\label{eq55}
8\pi(p - 3m_2\xi) = - \phi  m^2_{2}
\end{equation}
\begin{equation}
\label{eq56}
8\pi \rho = 3\phi m^2_{2}
\end{equation}
\noindent where, the Barber's scalar function $\phi$ is given by the Eq. (\ref{eq39}).\\
\bigskip
\noindent {\bf Model I $(\xi = \xi_0)$}\\
\noindent In this case, we assume $n = 0$. Further from Eq. (\ref{eq34}), we obtain 
$\xi$ = $\xi_O $= constant and hence Eq. (\ref{eq55}) reduces to
\begin{equation}
\label{eq57}
p = - \frac{\phi m^2_{2}}{8\pi} + 3m_2 \xi_0
\end{equation}
\bigskip
\noindent {\bf Model II $(\xi = \xi_0 \rho)$}\\
\noindent In this case, we assume $n = 1$. Further from Eq. (\ref{eq34}), we obtain 
$\xi$ = $\xi_O \rho$ and hence Eq. (\ref{eq55}) becomes
\begin{equation}
\label{eq58}
p = - \frac{\phi m^2_{2}}{8\pi}(1 - 9m_2 \xi_0)
\end{equation}
\section{Conclusion} 
In this paper we have obtained a class of LRS Bianchi Type I models in Barber's 
second self creation theory in the presence of bulk viscous fluid for constant 
deceleration parameter. Assuming an ad-hoc law for the coefficient of bulk viscosity 
of the form $\xi(t)$ = $\xi_o\rho^n$, where $\rho$ is the energy density and $n$
is the power index, we have obtained exact solutions of the field equations. The 
nature of the Barber's scalar function $\phi$ and energy density $\rho$ have been 
examined for values of $n$ = $0$ and $n$ = $1$ for both the (i) power-law and (ii) 
exponential expansion of both the non-flat and flat universe. The models discussed 
here are isotropic and homogeneous and, in view of the assumption of isotropy, the
shear viscosity is absent. The effect of bulk viscosity is to produce a change in 
the perfect fluid. We observe here that Murphy's conclusion about the absence of
a big bang type singularity in the finite past in models with bulk viscous fluid is, 
in general, not true.\\ 
\nonumsection{Acknowledgements} 
\noindent The authors would like to thank the Harish-Chandra Research Institute, Allahabad, 
India for providing  facility where this work was carried out. \\
\newline
\newline
\nonumsection{References}

\end{document}